\documentclass[a4paper]{article}

\usepackage{INTERSPEECH2022}
\usepackage{xcolor}
\usepackage{multirow,array,enumitem,adjustbox}
\usepackage{booktabs, threeparttable,caption, makecell, url}

\title{SJTU-AISPEECH System for VoxCeleb Speaker Recognition Challenge 2022}
\name{Zhengyang Chen$^{1,*}$, Bing Han$^{1,*}$, Xu Xiang$^{2,*}$, Houjun Huang$^{2,*}$, Bei Liu$^1$, Yanmin Qian$^{1}$ \thanks{\textsuperscript{*}Equal Contribution}}
\address{
  $^1$MoE Key Lab of Artificial Intelligence, AI Institute \\
  Department of Computer Science and Engineering, X-LANCE Lab, Shanghai Jiao Tong University \\
  $^2$AISpeech Ltd, Suzhou China}
\email{\{zhengyang.chen, hanbing97\}@sjtu.edu.cn, \{xu.xiang, houjun.huang\}@aispeech.com, \{beiliu, yanminqian\}@sjtu.edu.cn}

\begin{document}

\maketitle
\begin{abstract}
This report describes the SJTU-AISPEECH system for the Voxceleb Speaker Recognition Challenge 2022. For track1, we implemented two kinds of systems, the online system and the offline system. Different ResNet-based backbones and loss functions are explored. Our final fusion system achieved 3rd place in track1. For track3, we implemented statistic adaptation and jointly training based domain adaptation. In the jointly training based domain adaptation, we jointly trained the source and target domain dataset with different training objectives to do the domain adaptation. We explored two different training objectives for target domain data, self-supervised learning based angular proto-typical loss and semi-supervised learning based classification loss with estimated pseudo labels. Besides, we used the dynamic loss-gate and
label correction (DLG-LC) strategy to improve the quality of pseudo labels when the target domain objective is a classification loss. Our final fusion system achieved 4th place (very close to 3rd place, relatively less than 1\%) in track3.
\end{abstract}

\section{Track1: Close-set Supervised}
For track1, we implemented two kinds of systems in the challenge. The online system and offline system.

\subsection{Data Usage}
In the fully supervised speaker verification closed track, 5994 speakers with 1,092,009 utterances from Voxceleb2 dev set \cite{chung2018voxceleb2} are used as the training set.

\subsubsection{Offline System}
For the offline system, we did extensive data augmentation to generate more training data. 

\begin{itemize}
    \item Speed perturbation. We applied speed perturbation \cite{yamamoto2019speaker,wang2020dku,zhao2021speakin} by speeding up and slowing down the original audio with ratios 1.1 and 0.9 respectively. Since the pitch of the newly generated audio is changed, we consider it comes from a new speaker. Thus, we have 17,982 speakers and 3,276,027 utterances after speed perturbation augmentation.
    \item Additive noises and reverberation. Following kaldi voxceleb \footnote{\url{https://github.com/kaldi-asr/kaldi/tree/master/egs/voxceleb/v2}} recipe, we generate another four copies of training data using the noise and reverberation samples from MUSAN \cite{musan2015} and RIR NOISES \footnote{\url{https://www.openslr.org/28}}.
    
\end{itemize}
After offline data augmentation, we totally have 15 times data compared to the original Voxceleb dev set. Then, we extracted 80-dimentional fbank feature for all the data and no voice activity detection (VAD) is applied. Besides, we applied 300-frame sliding window mean normalization on the features before feeding them to the neural network.

\subsubsection{Online System}
\label{sssec:track1_data_online}
Different from the offline system, we perform online data augmentation and feature extraction for the online systems. For each sampled audio in the training process, we process it for five steps:
\begin{enumerate}
    \item Randomly sample a speed perturbation ratio from $\{1.0, 1.1, 0.9\}$ and do speed perturbation augmentation (ratio 1.0 means no speed perturbation).
    \item Decide whether to do noise augmentation with probability 0.6. If doing noise augmentation, randomly sampled a noise type from $\{\text{babble}, \text{noise}, \text{music}, \text{reverb}\}$. These four types of noise are the same as the noise in the offline data processing.
    \item Randomly sample a segment from the processed audio with a pre-defined segment length.
    \item Extract 80-dimensional fbank feature from audio segment.
    \item Do mean normalization on the fbank feature.
    
\end{enumerate}

\subsection{System Architectures}

\begin{table}[ht!]
\centering
\caption{\textbf{Track1 system architecture configuration.} We use c32 to denote ResNet with base channel 32 and use c64 to denote ResNet with base channel 64.}
\begin{adjustbox}{width=.48\textwidth,center}
\begin{threeparttable}
\begin{tabular}{lllc}
\toprule
Index & Backbone & Pooling Method  & Param \# \\
\hline
Online System & & & \\
\hline
S1  & ResNet34-c64 & MQMHA &  27.8M \\
S2  & ResNet101-c32 & MQMHA & 23.8M \\
S3  & ResNet101-c64 & MQMHA & 68.7M \\
S4  & ResNet152-c32 & MQMHA & 27.7M \\
S5  & ResNet221-c32 & MQMHA & 31.6M \\
\hline
Offline System & & & \\
\hline
S6  & ResNet152-c32 & Statistic Pooling & 19.7M \\
S7  & Res2Net101-c32 & Statistic Pooling & 28.6M \\
\bottomrule
\end{tabular}
\end{threeparttable}
\label{table:track1_systems}
\end{adjustbox}
\end{table}
We list all our systems' configuration for track1 in Table \ref{table:track1_systems}.
In this challenge, we mainly used the ResNet \cite{he2016deep,zeinali2019but} and Res2Net \cite{gao2019res2net,xiang2020xx205} based architecture for speaker verification system. Two different configurations for the base channel number of ResNet are tried, channel number 32 (c-32) and channel number 64 (c-64). The classical statistic pooling layer \cite{snyder2017deep} is used in offline systems and multi-query multi-head attention pooling mechanism (MQMHA) proposed in \cite{zhao2021speakin} is used in online systems.




\subsection{System Training}
For both online and offline system, we train the system for two stages including pre-training and large margin fine-tuning. And all our systems are implemented based on WeSpeaker\footnote{\url{https://github.com/wenet-e2e/wespeaker}} toolkit.

\subsubsection{Stage I: Pre-training}
    In the stage I, we sample 2s segment from each utterance to construct the training batch. SGD is used as the optimizer with initial learning rate 0.1 and final learning rate 0.00005. The learning rate exponentially decreases from initial value to the final value during the training process. We set the weight decay to 1e-4 for the online system and 1e-3 for the offline system. Besides, we train the online system using AAM \cite{xiang2019margin,deng2019arcface} + K-subcenter \cite{deng2020sub} + Inter-TopK \cite{zhao2021speakin} loss. We set the scale and margin in AAM loss to 32.0 and 0.2, set the sub-center number K in K-subcenter loss to 3 and set the extra penalty and topK in Inter-TopK loss to 0.06 and 5 respectively. we train the offline system use AM ~\cite{wang2018additive,wang2018cosface} + K-subcenter \cite{deng2020sub} loss. The K-subcenter number in the offline system is 3. The margin and scale for AM loss are set to 0.2 and 32.0 respectively.

\subsubsection{Stage II: Large Margin Fine-tuning}
In stage II, we do the large margin fine-tuning \cite{thienpondt2021idlab} for online and offline systems. In this stage, speed perturbation augmentation and Inter-TopK loss are removed. All systems are finetuned with 6s training segments. For online systems, the margin for AAM loss is increased from 0.2 to 0.5, while for offline systems, the margin for AM loss is increased from 0.2 to 0.35.


\subsection{System Evaluation}
\label{ssec:track1_eval}
For both online and offline systems, cosine distance is used to measure the similarity between enrollment and test utterance. We also do the adaptive score normalization (as-norm) \cite{cumani2011comparison} and set the imposter cohort size to 600. The imposter cohort is estimated from the original voxceleb2 dev set by averaging the embeddings for each training speaker. Besides, we construct a trial with 30k pairs from voxceleb2 dev set following the strategy proposed in \cite{thienpondt2021idlab}  to do score calibration.

\subsection{Results}
We list the results for track1 systems in Table \ref{table:track1_res}. For the online system, deeper ResNet seems to have better performance and the ResNet221 achieves the best result. Compared with the online system, the offline system has better minDCF when the EER is comparable. Finally, we fused the scores of all these seven systems to get the final submission system.

\begin{table}[ht!]
\centering
\caption{\textbf{System results on track1 validation set and test set.} Without specific notation, the results are evaluated on the validation set.}
\begin{adjustbox}{width=.45\textwidth,center}
\begin{tabular}{llll}
\hline 
System Type & System Index & EER (\%)  & minDCF \\
\hline
\multirow{5}{*}{Online} & S1 & 2.001 & 0.1375 \\
& S2 & 1.551 & 0.1029  \\
& S3 & 1.566 & 0.1034 \\
& S4 & 1.457 & 0.1036 \\
& S5 & 1.420 & 0.0976 \\
\hline
\multirow{2}{*}{Offline} & S6 & 1.609 & 0.0980 \\

& S7 & 1.480 & 0.0890 \\
\hline

Fusion & - & 1.056 & 0.0686 \\
Fusion (test set) & - & 1.911 & 0.1010 \\
\hline

\end{tabular}
\label{table:track1_res}
\end{adjustbox}
\end{table}

\section{Track3: Semi-Supervised}
In this track, we only implement the online system and we also use the data processing strategy introduced in section \ref{sssec:track1_data_online}.

\subsection{Data Usage}
\begin{table}[ht!]
\centering
\caption{\textbf{Training set for Track3.}}
\begin{adjustbox}{width=.45\textwidth,center}
\begin{tabular}{lllll}
\hline 
Domain & Data Source & Speaker \# & Utt \# & Duration \\
\hline
Source & Voxceleb2 dev & 5,994 & 1,092,009 & 2,360 hrs \\
Target & Cnceleb2 & N/A & 454,839 & 970 hrs \\
Target & Cnceleb & 50 & 1,000 & 2.57 hrs \\
\hline
\end{tabular}
\label{table:track3_training_set}
\end{adjustbox}
\end{table}

In this semi-supervised domain adaptation closed track, the training set contains three parts, the large-scale labeled source domain dataset from Voxceleb2 dev set \cite{chung2018voxceleb2}, the unlabeled large-scale target domain dataset from Cnceleb2 \cite{li2022cn} and the small amount of labeled dataset from Cnceleb \cite{fan2020cn}. We listed the statistics of these three datasets in Table \ref{table:track3_training_set}. We will leverage different training sets in Table \ref{table:track3_training_set} for the different adaptation methods proposed in section \ref{ssec:domain_adaptation}.


\subsection{Domain Adaptation Strategy}
\label{ssec:domain_adaptation}

\subsubsection{Statistic Adaptation}
\label{sssec:statistic_adaptation}
During the evaluation process, the extracted embedding is always mean-normalized by the statistics derived from a specific dataset and we usually use the statistics of the training set to do this normalization. However, when the training set has a large domain mismatch with the evaluation set, the statistic may not be optimal. Because we have the unsupervised data from the target domain in this track, we can derive the statistic from this dataset and we call this method as statistic adaptation (SA) in our experiment. Based on the statistic domain adaptation, we can directly evaluate the online system from track1 on the track3 evaluation set.

\begin{figure}[ht]
  \centering
  \includegraphics[width=0.6\linewidth]{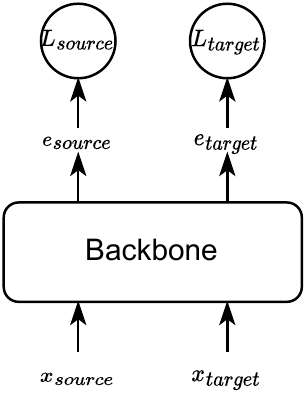}
  \caption{Jointly training based domain adaptation for track3.}
  \label{fig:source_target_joint_train}
  \vspace{-0.5cm}
\end{figure}
\subsubsection{Jointly Training Based Domain Adaptation}
As shown in Figure \ref{fig:source_target_joint_train}, here we propose to jointly train the source and target domain data to do domain adaptation. We only used the target domain unlabeled data in this training process. In the training process, we sample equal number source and target domain samples to construct each training batch. 

The training objective for the jointly training is defined as:
\begin{equation}
    \mathcal{L} = \mathcal{L}_{source} + \mathcal{L}_{target}
\end{equation}
where $\mathcal{L}_{source}$ and $\mathcal{L}_{target}$ are the losses for source and target domain data respectively. The stage I pre-training model from track1 is used to initialize the jointly training model and the $\mathcal{L}_{source}$ is the same as the original loss used in the pre-training model. In this challenge, we explored three different objectives to calculate the $\mathcal{L}_{target}$:
\begin{enumerate}
    \item Angular prototypical loss (APL): The same self-supervised learning based angular prototypical loss with \cite{chen2021self} is used to do domain adaptation. There is no need to know each utterance's speaker indentity. Positive pairs are sampled from the same utterance and negative pairs are sampled from different utterances.
    \item Two-head classification loss (TCL): Here, we use $k$-means to estimate the pseudo labels for the target domain unsupervised data based on a specific well-trained model. Then, another AAM-softmax based classification head is added to calculate $\mathcal{L}_{target}$.
    \item One-head classification loss (OCL): Similar to TCL, we first estimate the pseudo labels. The difference is that the target domain data shared the same classification head with the source domain data. 
\end{enumerate}

\subsubsection{Dynamic Loss-gate and Label Correction}

Besides, the estimated pseudo labels from a pre-trained model may be inaccurate. Such low-quality labels will confuse and degrade the model during the training, which indicates that finding a way to select high-quality labels is the key to improving performance. To filter out the samples with unreliable labels, we adopted the dynamic loss-gate and label correction (DLG-LC) which is proposed in our previous work \cite{han2022self}. By modeling the loss histogram with Gaussian distribution, we can obtain a dynamic loss-gate to select reliable data when training on pseudo labels. Instead of dropping unreliable data, we incorporate a predicted posterior probability as the target distribution to prevent over-fitting into incorrect samples. 

\subsection{System Training}
In this domain adaptation scenario, we use the track1 online system after stage I training as the pre-training model. We either directly evaluate it on the target domain or use it as an initial model to do jointly fine-tuning on the source and target domain data. For the jointly fine-tuning, equal number of utterances from source domain data set and unlabeled target domain data set are sampled to construct each training batch. We also use the 2 seconds segment in the training process. In this fine-tuning stage, the initial learning rate and final learning rate are decreased to 0.001 and 0.00001 respectively.

\subsection{System Evaluation}
\label{ssec:track3_eval}
Similar to the scoring strategy in section \ref{ssec:track1_eval}, cosine similarity, adaptive score normalization, and score calibration are applied to score the trial. The imposter cohort is estimated from the target domain unsupervised dataset. Because we don't know the speaker label for this unsupervised dataset, we use the pseudo speaker label in imposter cohort estimation. Besides, we construct a validation trial with 19k pairs from the track3 supervised dataset to do score calibration.

\subsection{Results}
\begin{table}[ht!]
\centering
\caption{\textbf{System results on track3 validation set with different back-end processing method.} Here, we directly evaluate the online system ResNet34-c64 from track1 on the track3 validation set. Despite that we cannot use the speaker label of track3 unsupervised set, we still list the results of as-norm when using the ground truth speaker label in the bracket for comparison. The pseudo labels for as-norm are estimated from the APL system in Table \ref{table:track3_res}.}
\begin{adjustbox}{width=.45\textwidth,center}
\begin{tabular}{lll}
\hline 

Back-end Processing Method & EER (\%) & minDCF \\ \hline
ResNet34-c64 & & \\
cosine scoring & 14.60 & 0.5302 \\
+ statistic adaptation & 11.65 &	0.4552 \\
++ as-norm & 11.58 (11.735) & 0.4032 (0.3959) \\
+++ score-calibration & 9.950 & 0.4290 \\

\hline

\end{tabular}
\label{table:track3_backend_ablation}
\end{adjustbox}
\end{table}

We first evaluate the effect of different back-end processing methods introduced in section \ref{ssec:track3_eval} and list the results in Table \ref{table:track3_backend_ablation}. The statistic adaptation method in section \ref{sssec:statistic_adaptation} can also be considered as a kind of back-end processing method and we also involve it in this study. From the results, we find that the statistic adaptation strategy improves performance significantly. Then, we compare the as-norm results when using the ground truth speaker label or the pseudo label. It is interesting that whether the speaker label is accurate or not has little effect on the final results. Besides, the as-norm operation improves the minDCF metric a lot but has little effect on the EER metric or eval degrades the EER. Finally, we do the score calibration. The score calibration can improve the EER by a large margin but degrades minDCF metric. Because EER is the main metric for track3, we use all these back-end processing methods in the following experiments.

\begin{table}[ht!]
\centering
\caption{\textbf{System results on track3 validation set and test set.} Without specific notation, the results are evaluated on the validation set. The pseudo labels used for track3 systems are estimated from the APL system (the third line of the table).}
\begin{adjustbox}{width=.45\textwidth,center}
\begin{tabular}{llll}
\hline 
Adaptation Method & Backbone & EER (\%) & minDCF \\
\hline
Statistic Adaptation & ResNet34-c64 & 9.950 & 0.4290 \\
\hline
\multirow{1}{*}{APL} & ResNet34-c64 & 9.050 & 0.4063 \\

\multirow{1}{*}{TCL} & ResNet34-c64 & 9.320 & 0.4254 \\ \hline
\multirow{1}{*}{TCL + DLG-LC} & ResNet34-c64 & 9.060 & 0.4180 \\
\hline
\multirow{5}{*}{OCL} & ResNet34-c64 & 8.825 & 0.4104 \\
& ResNet101-c32 & 8.255 & 0.3771 \\
& ResNet152-c32 & 8.095 & 0.3696 \\
& ResNet101-c64 & 8.025 & 0.3670 \\
& ResNet221-c32 & 8.255 & 0.3674 \\
\hline
\multirow{1}{*}{OCL + DLG-LC} & ResNet152-c32 & 7.855 & 0.3681 \\
\hline

Fusion &  - & 7.135 & 0.3290 \\
Fusion (test set) & - & 8.087 & 0.4370 \\
\hline

\end{tabular}
\label{table:track3_res}
\end{adjustbox}
\end{table}

Then, we compare all the adaptation methods and list the results in Table
\ref{table:track3_res}. From the results, we find that all the jointly training based methods can improve the system's performance further. It is worth noting that the TCL strategy using estimated pseudo labels even performs worse than the APL strategy based on self-supervised learning. Encouragingly, the DLG-LC can enhance the TCL strategy to enable it to have comparable performance with APL, which indicates that it is important to filter out the noisy labels in TCL fine-tuning strategy. Besides, we find that it is better to use one classification head than two classification heads when doing the source and target domain jointly fine-tuning, which is reasonable.  When using two classification heads in TCL strategy, the speaker embedding only needs to be far away from the different speaker embeddings from the same domain. However, when there is only one classification head in OCL strategy, the speaker embedding should be far away from the different speaker embeddings from both source and target domains, which will make the network more discriminative. Besides, the DLG-LC also benefits the OCL strategy and can make further improvements. Finally, we fuse all the systems in Table \ref{table:track3_res} to get our submission system.

\section{Conclusion}
In this report, we give a detailed description of our submission to Track 1 \& 3 of VoxSRC 2022. In Track 1, different ResNet-based backbones and loss functions with online or offline features are explored. In Track 3, we investigate several training objectives for target domain data including angular proto-typical loss and classification loss. With DLG-LC, the performance is improved further. Finally, our fusion system achieved $3^{rd}$ and $4^{th}$ place in Track 1 and 3 respectively.



\bibliographystyle{IEEEtran}

\bibliography{mybib}

\end{document}